# A 20-kV Optical-Injection-Enhanced 4H-SiC Reverse-Conducting IGBT With Snapback-Free Operation and Low On-State Voltage

Yujian Chen, Guoliang Zhang, Ruijun Zhang, Liangxun Yue, Zhanwei Shen, Feng Zhang, Member, IEEE, and Rong Zhang, Member, IEEE

***Abstract*—This letter proposes a 20-kV 4H-SiC optical-injection-enhanced reverse-conducting IGBT (OIE-RC-IGBT) with snapback-free operation and low on-state voltage. Light-emitting diodes (LEDs) are integrated into the gate structure to provide optical injection during forward conduction. The emitted light generates a high concentration of photogenerated carriers in the carrier-storage layer (CSL) and drift region, thereby reducing the resistance of the JFET and drift regions and increasing the device current. Consequently, the collector-side $P^+$/N junction is turned on earlier, effectively suppressing the snapback effect and facilitating the transition to bipolar conduction. After bipolar conduction is established, the photogenerated carriers further enhance conductivity modulation, resulting in a reduced on-state voltage. TCAD simulation results show that, compared with the conventional RC-IGBT (C-RC-IGBT), the proposed device achieves snapback-free operation while reducing the required number of MOS cells from 25 to 5. Moreover, the on-state voltage, turn-off fall time, and turn-off energy loss are reduced by 22.37%, 25.06%, and 19.69%, respectively. These results demonstrate the potential of optical injection for improving the conduction and turn-off performance of ultrahigh-voltage SiC RC-IGBTs.**

## I. Introduction

4H-SiC is a promising material for high-voltage power devices owing to its high critical electric field, high saturation carrier velocity, and high thermal conductivity [1], [2], [3]. SiC MOSFETs offer fast switching; however, their on-state voltage increases significantly at voltage ratings above 10 kV [4], [5]. In contrast, SiC IGBTs exploit conductivity modulation to achieve low conduction loss, making them attractive for ultrahigh-voltage applications [6], [7].

Conventional IGBTs require antiparallel freewheeling diodes (FWDs) for reverse conduction, resulting in additional parasitic effects and power loss [8], [9]. Reverse-conducting IGBTs (RC-IGBTs) monolithically integrate the IGBT and FWD functions into a single chip, enabling bidirectional conduction. However, conventional RC-IGBTs suffer from the snapback effect during forward conduction. Existing approaches generally modify the collector or drift region to increase the collector-short resistance and suppress snapback, often at the expense of fabrication complexity or electrical performance [10], [11], [12], [13], [14], [15], [16].

In this letter, a 20-kV 4H-SiC optical-injection-enhanced RC-IGBT (OIE-RC-IGBT) with gate-integrated LEDs is proposed. During forward conduction, the emitted UV light generates additional photogenerated carriers, increasing the device current and facilitating the turn-on of the collector-side P+/N junction, thereby suppressing snapback. After bipolar conduction is established, the photogenerated carriers further enhance conductivity modulation and reduce the on-state voltage [17]. The device mechanism and characteristics are investigated using TCAD simulations.

This research was supported by the National Natural Science Foundation of China (Grant Nos. 62274137, U25A20492), State Key Laboratory of Electrical Insulation and Power Equipment (EIPE26212), National Key Research and Development Program of China (Grant No. 2023YFB3609500, Grant No. 2023YFB3609502).

Y. Chen, G. Zhang, L. Yue, F. Zhang, and Rong Zhang are with the College of Physical Science and Technology, Xiamen University, Xiamen 361005, China. F. Zhang and Rong Zhang are also with the School of Integrated Circuits, Xiamen University, Xiamen 361005, China. Ruijun Zhang is with the School of Information Engineering, Xiamen Ocean Vocational College, Xiamen 361005, China. Z. Shen is with the Key Laboratory of Semiconductor Material Sciences, Institute of Semiconductors, Chinese Academy of Sciences, Beijing 100083, China (corresponding authors: Feng Zhang and Rong Zhang; e-mail: fzhang@xmu.edu.cn; rzhangxmu@xmu.edu.cn).

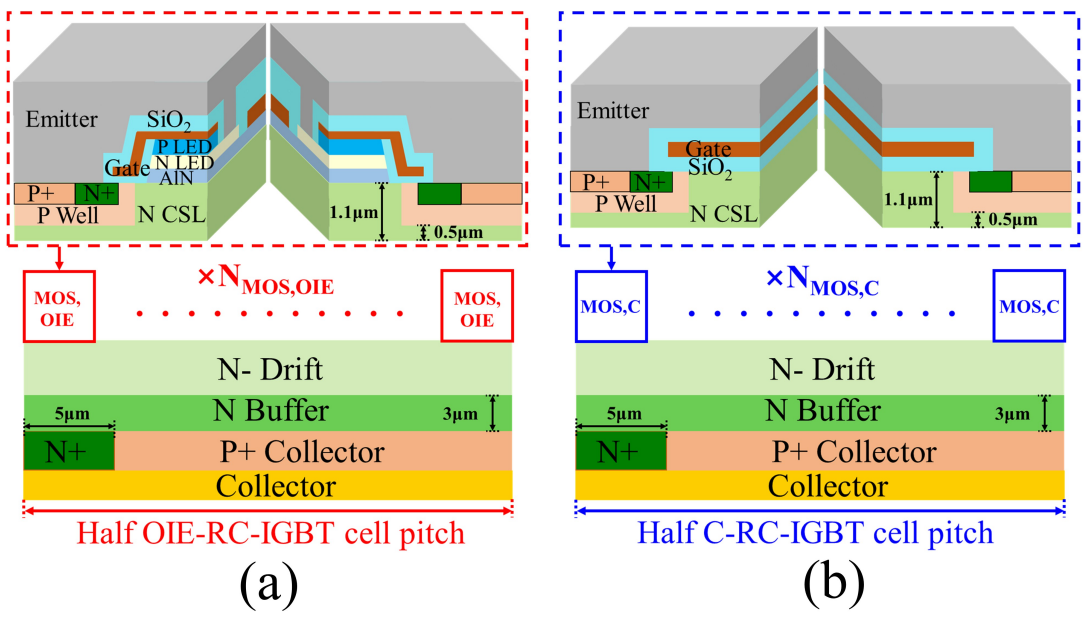

Fig. 1. Schematic structures of (a) the OIE-RC-IGBT and (b) the C-RC-IGBT.

## II. Device Structure and Mechanism

Fig. 1(a) shows a schematic cross section of the proposed OIE-RC-IGBT. Compared with the C-RC-IGBT in Fig. 1(b),

TABLE I
MAJOR DEVICE PARAMETERS

| Parameters | OIE-RC-IGBT | C-RC-IGBT |
|---|---|---|
| Gate oxide thickness, $t_{OX}$(nm) | 50 | 50 |
| AlN thickness, $t_{AlN}$(nm) | 100 | --- |
| Channel length, $L_{ch}$ (μm) | 0.5 | 0.5 |
| JFET region length, $L_j$ (μm) | 10 | 10 |
| MOS cell pitch, $L_{MOS}$ (μm) | 25 | 25 |
| N⁺ collector length, $L_{cn}$ (μm) | 5 | 5 |
| LED length, $L_{LED}$ (μm) | 8 | --- |
| CSL doping, $N_{CSL}$ (cm$^{-3}$) | $5\times10^{15}$ | $5\times10^{15}$ |
| N buffer doping, $N_{bu}$ (cm$^{-3}$) | $2\times10^{16}$ | $2\times10^{16}$ |
| N⁺ collector doping, $N_{cn}$ (cm$^{-3}$) | $1\times10^{19}$ | $1\times10^{19}$ |
| P⁺ collector doping, $N_{cp}$ (cm$^{-3}$) | $1\times10^{20}$ | $1\times10^{20}$ |

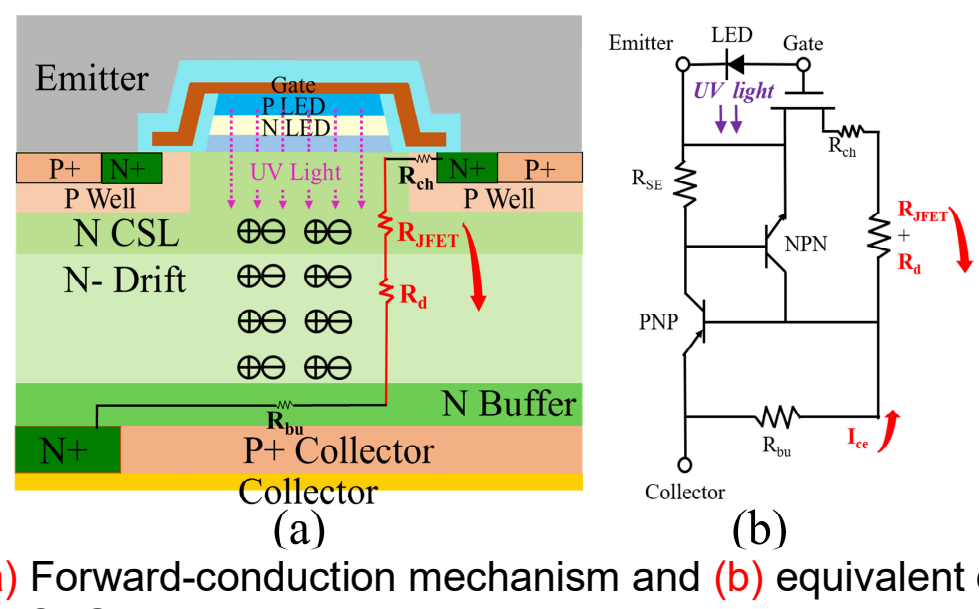

Fig. 2. (a) Forward-conduction mechanism and (b) equivalent circuit of the OIE-RC-IGBT.

the proposed device additionally integrates LEDs into the gate structure. Major device parameters are summarized in Table I.

Fig. 2 illustrates the forward-conduction mechanism and equivalent circuit of the OIE-RC-IGBT. Under positive gate and collector biases, the integrated LEDs emit UV light that generates photogenerated carriers, reducing $R_{JFET}$ and $R_d$ and thereby increasing the total device current ($I_{ce}$). Before the collector junction is fully turned on, conduction is dominated by the MOSFET path. Following the voltage-drop criterion for collector-junction turn-on in [18], the junction turns on when $I_{ce}\times R_{bu} \geq V_{bi}$. Using the ideal abrupt-junction approximation and the doping concentrations in Table I, $V_{bi}$ is analytically estimated to be approximately 3.12 V at 300 K. Thus, optical injection facilitates collector-junction turn-on and suppresses snapback.

## III. RESULTS AND DISCUSSION

The TCAD simulations include doping-dependent mobility, high-field saturation, Shockley–Read–Hall (SRH) and Auger recombination, incomplete ionization, and Okuto–Crowell impact ionization. The optical model employs a monochromatic light source, with AlN absorption and interface reflection taken into account [19], [20]. Both devices employ a 200-μm drift region doped at 2 × $10^{14}$ cm$^{-3}$, providing a simulated blocking capability of 20 kV. The integrated LED is modeled with λ = 370 nm, $I_{opt}$ = 25 W/cm², and a threshold voltage of 4 V [21]. The minority-carrier lifetime is set to 5 μs, consistent with experimental values [22], [23], [24]. Other device parameters are listed in Table I.

Fig. 3(a) shows the forward and blocking characteristics. Here, $\Delta V_{SB} = V_{SB} - V_H$ is defined as the snapback voltage,

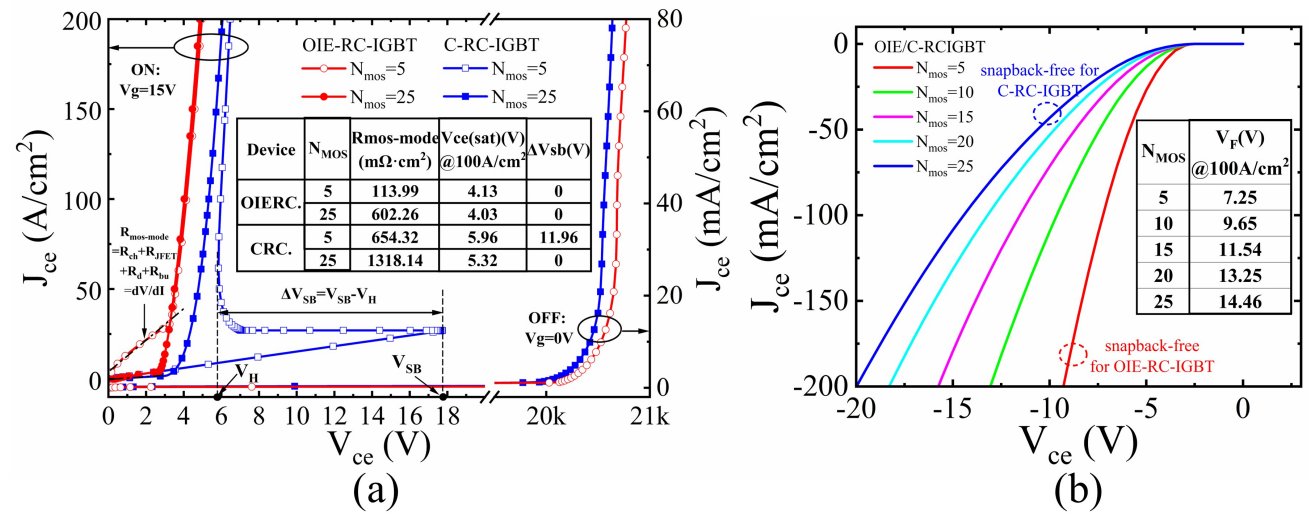

Fig. 3. (a) Forward conduction and blocking characteristics and (b) reverse conduction characteristics of the proposed and conventional RC-IGBTs.

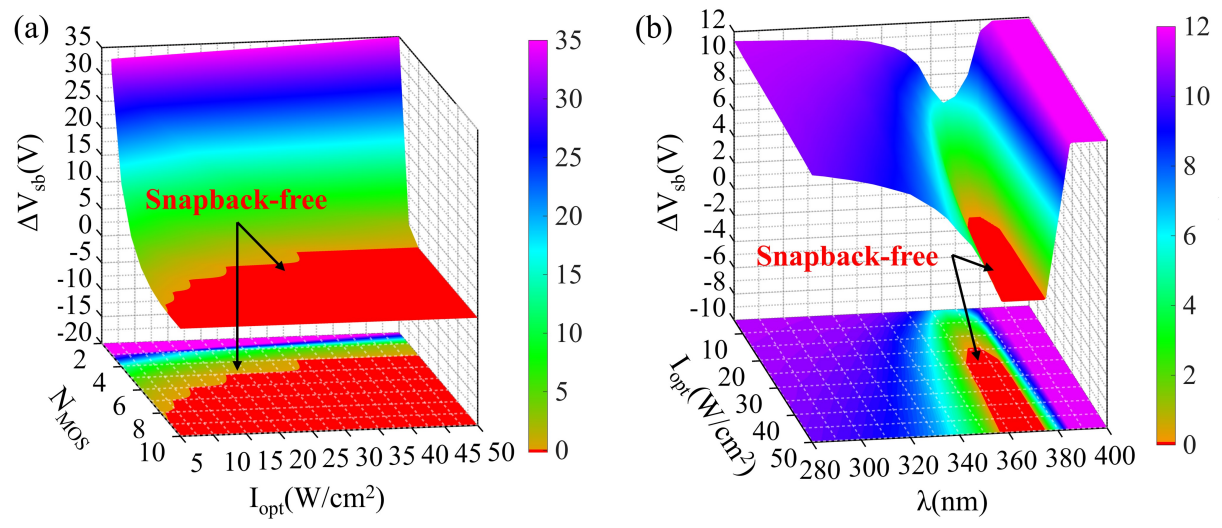

Fig. 4. Dependence of $\Delta V_{SB}$ on (a) $I_{opt}$ and $N_{MOS}$ and (b) $I_{opt}$ and λ for the OIE-RC-IGBT.

where $V_{SB}$ is the snapback turnover voltage and $V_H$ is the minimum collector voltage after snapback [12]. Under 370-nm UV illumination at 25 W/cm², the OIE-RC-IGBT achieves snapback-free operation ($\Delta V_{SB}$ = 0) with an 80% reduction in RC-IGBT cell pitch, corresponding to a reduction in the number of MOS cells ($N_{MOS}$) from 25 to 5. Optical injection also enhances conductivity modulation [17]; at 100 A/cm², the OIE-RC-IGBT ($N_{MOS}$ = 5) exhibits $V_{ce(sat)}$ = 4.13 V, 22.37% lower than the 5.32 V of the C-RC-IGBT ($N_{MOS}$ = 25). As shown in Fig. 3(b), both devices exhibit identical reverse-conduction characteristics at the same NMOS. Under snapback-free conditions, the OIE-RC-IGBT ($N_{MOS}$ = 5) exhibits $V_F$ = 7.25 V at 100 A/cm², 49.86% lower than the 14.46 V of the C-RC-IGBT ($N_{MOS}$ = 25).

Fig. 4 shows the effects of $N_{MOS}$, optical intensity ($I_{opt}$), and wavelength (λ) on $\Delta V_{SB}$ of the OIE-RC-IGBT. As shown in Fig. 4(a), $\Delta V_{SB}$ decreases with increasing $I_{opt}$ and $N_{MOS}$. Even at $I_{opt}$ = 5 W/cm², snapback-free operation is achieved for $N_{MOS} \geq 9$, indicating effective snapback suppression at relatively low optical intensity. Fig. 4(b) shows that the strongest suppression occurs at 360–375 nm. At shorter wavelengths, photons are absorbed closer to the surface, limiting the reduction in $R_{JFET}$ and $R_d$ [17], whereas longer wavelengths near the 4H-SiC absorption edge result in insufficient optical absorption and photogeneration, thereby weakening snapback suppression [25], [26].

Fig. 5 compares the turn-off characteristics of the snapback-free OIE-RC-IGBT ($N_{MOS}$ = 5, $I_{opt}$ = 25 W/cm², λ = 370 nm) and C-RC-IGBT ($N_{MOS}$ = 25) using the simulation circuit shown in Fig. 5(a). As shown in Fig. 5(b), the OIE-RC-IGBT exhibits a fall time ($t_f$) of 670.3 ns, 25.06% shorter than the 894.5 ns of the C-RC-IGBT. $E_{off}$, calculated by integrating the instantaneous power over the defined $t_f$ interval, is reduced by 19.69%, from 193.12 to 155.1 mJ/cm². The improved turn-off performance is attributed to the larger

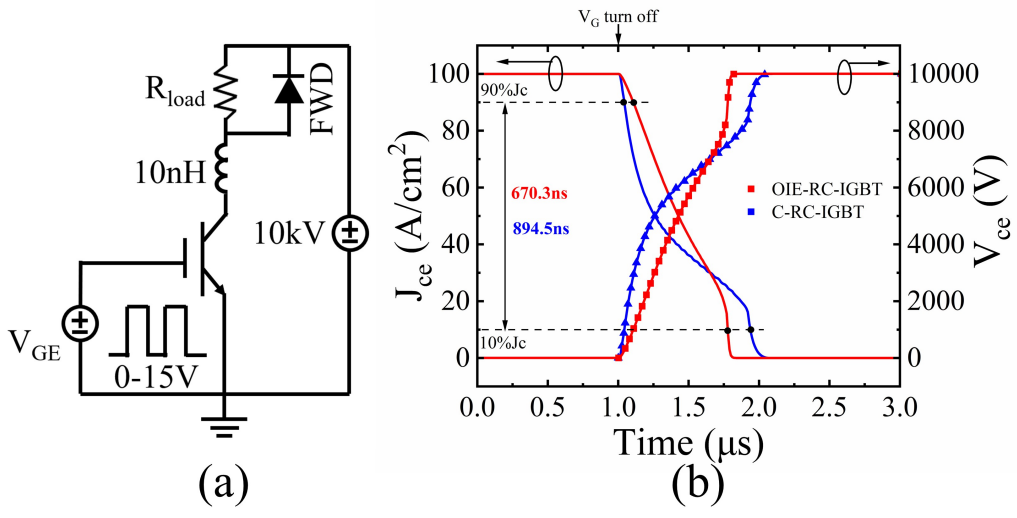

Fig. 5. (a) Simulation circuit for turn-off analysis and (b) turn-off waveforms of the snapback-free OIE-RC-IGBT ($N_{MOS}$ = 5, $I_{opt}$ = 25 W/cm², λ = 370 nm) and C-RC-IGBT ($N_{MOS}$ = 25).

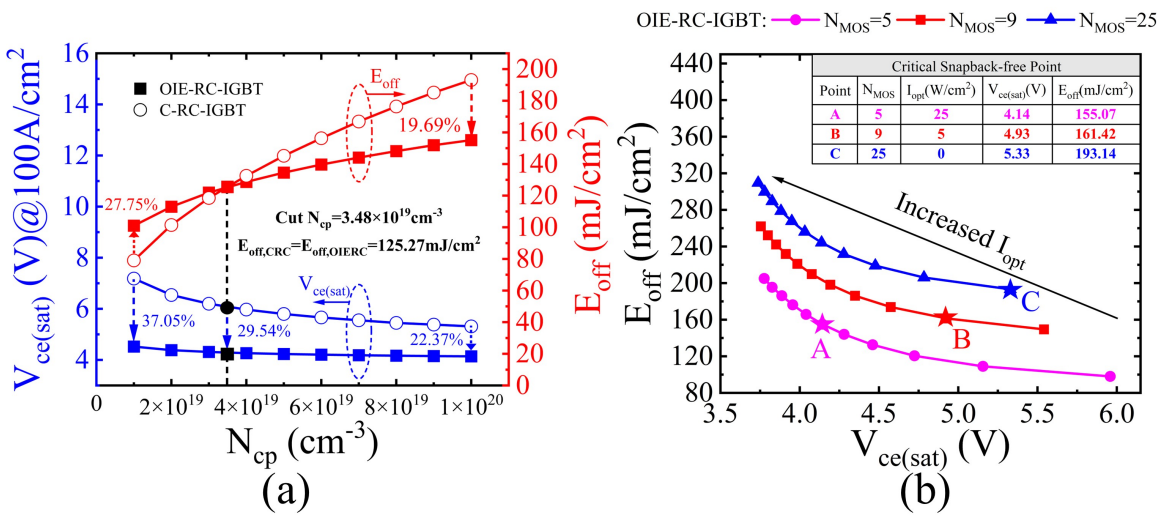

Fig. 6. (a) Effect of $P^+$ collector doping $N_{cp}$ on the $V_{ce(sat)}$–$E_{off}$ trade-off for OIE-RC-IGBT ($N_{MOS}$ = 5, $I_{opt}$ = 25 W/cm², λ = 370 nm) and C-RC-IGBT ($N_{MOS}$ = 25). (b) Effect of $I_{opt}$ and $N_{MOS}$ on the $V_{ce(sat)}$–$E_{off}$ trade-off for OIE-RC-IGBT (λ = 370 nm).

fraction of the unipolar N+ collector region, which facilitates carrier extraction.

Fig. 6(a) compares the $V_{ce(sat)}$–$E_{off}$ tradeoff versus $N_{cp}$. The OIE-RC-IGBT is less sensitive to $N_{cp}$ because photogenerated carriers compensate for variations in P+-collector injection. For $N_{cp} > 3.48 \times 10^{19}$ cm⁻³, enhanced carrier injection and optical injection reduce $V_{ce(sat)}$, while the larger N+ collector fraction facilitates carrier extraction and reduces Eoff. For $N_{cp} < 3.48 \times 10^{19}$ cm⁻³, the C-RC-IGBT exhibits lower $E_{off}$ but higher $V_{ce(sat)}$. At $N_{cp} = 3.48 \times 10^{19}$ cm⁻³, both devices exhibit Eoff = 125.27 mJ/cm², while the OIE-RC-IGBT achieves $V_{ce(sat)}$ = 4.28 V, 29.54% lower than the C-RC-IGBT. Fig. 6(b) shows that increasing Iopt reduces $V_{ce(sat)}$ at the expense of increased Eoff. At the snapback-free threshold, smaller $N_{MOS}$ provides an improved tradeoff owing to the larger N+ collector fraction.

Fig. 7 outlines the fabrication process of the OIE-RC-IGBT. The N+/P+ collectors and active cells are first formed by reverse epitaxy and ion implantation [27], [28]. AlN deposition, LED growth, P-LED etching, mesa formation, and gate-oxide deposition are then performed [17], [29]. After opening the oxide, polysilicon is patterned to form the LED anode and IGBT gate, followed by interlayer dielectric deposition, contact formation, and metallization. During forward conduction, positive gate and collector biases activate the LED for UV optical injection. Alternatively, an external laser/LED can provide optical injection through windows in the metal pad.

## IV. Conclusion

This letter proposes an optical-injection approach for achieving snapback-free operation and low $V_{ce(sat)}$ in RC-IGBTs. UV light emitted by the integrated LEDs generates photocarriers during forward conduction, suppressing snapback and enhancing conductivity modulation with a reduced

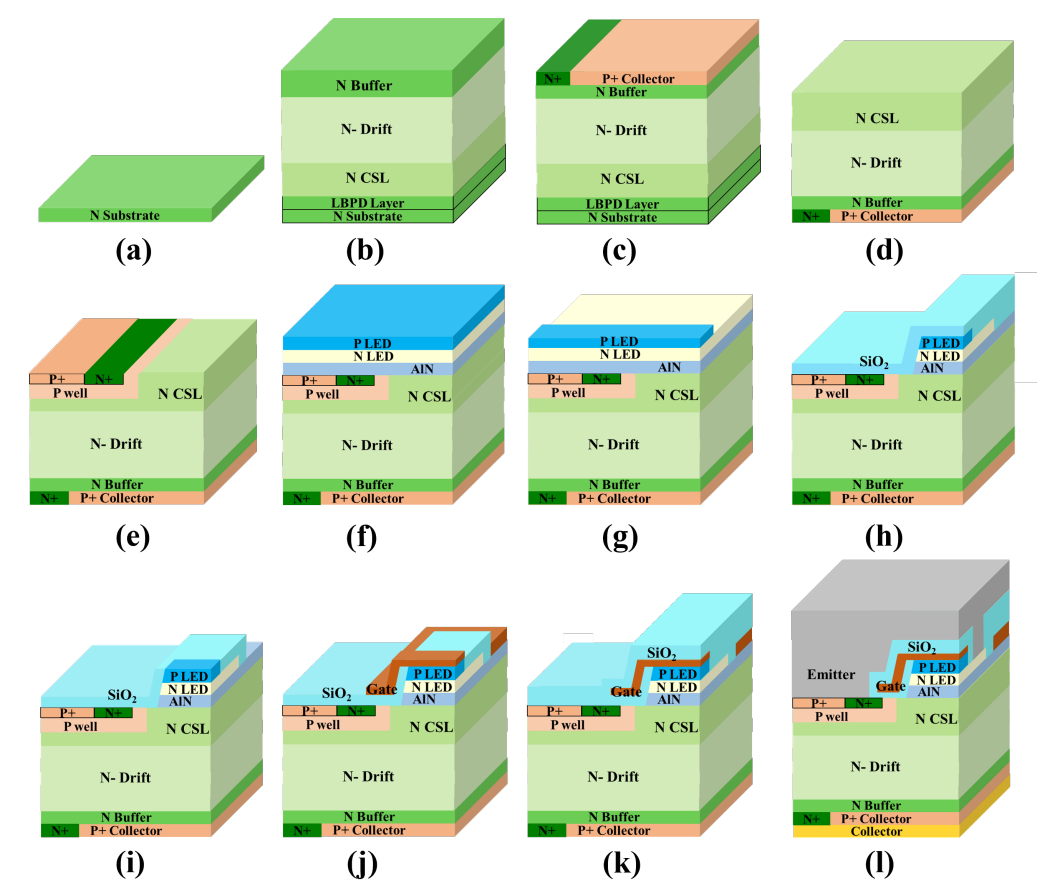

Fig. 7. Key fabrication steps of the OIE-RC-IGBT.

RC-IGBT cell pitch. The reduced cell pitch further improves reverse-conduction and turn-off performance, demonstrating the potential of the proposed device for ultrahigh-voltage applications.